\documentclass[twocolumn,pra,aps,superscriptaddress]{revtex4-2}
\usepackage{mathtools}
\usepackage{soul}
\usepackage{amsbsy}
\usepackage{amstext}
\usepackage[unicode=true,pdfusetitle,
 bookmarks=false,
 breaklinks=false,pdfborder={0 0 1},backref=false,colorlinks=false]
 {hyperref}
\usepackage[dvipsnames]{xcolor}
 \hypersetup{colorlinks=true,citecolor=NavyBlue,filecolor=black,linkcolor=black,urlcolor=black}

\makeatletter

\@ifundefined{textcolor}{}
{%
 \definecolor{BLACK}{gray}{0}
 \definecolor{WHITE}{gray}{1}
 \definecolor{RED}{rgb}{1,0,0}
 \definecolor{GREEN}{rgb}{0,1,0}
 \definecolor{BLUE}{rgb}{0,0,1}
 \definecolor{CYAN}{cmyk}{1,0,0,0}
 \definecolor{MAGENTA}{cmyk}{0,1,0,0}
 \definecolor{YELLOW}{cmyk}{0,0,1,0}
}

\newcommand\ket[1]{\left|#1\right\rangle}
\newcommand\bra[1]{\left\langle #1 \right|}

\usepackage{amsmath}
\usepackage{graphicx}
\usepackage{amssymb}
\usepackage{txfonts,color}
\usepackage{dsfont}
\makeatother

\begin{document}
\title{Deterministic multipartite entanglement via fractional state transfer across quantum networks}
\author{G. F. Pe{\~n}as}
\affiliation{Instituto de F{\'i}sica Fundamental (IFF), CSIC, Calle Serrano 113b, 28006 Madrid, Spain}
\author{J.-J. Garc{\'i}a-Ripoll}
\affiliation{Instituto de F{\'i}sica Fundamental (IFF), CSIC, Calle Serrano 113b, 28006 Madrid, Spain}
\author{R. Puebla}
\affiliation{Departamento de F{\'i}sica, Universidad Carlos III de Madrid, Avda. de la Universidad 30, 28911 Legan{\'e}s, Spain}

\begin{abstract}
 The generation of entanglement across different nodes in distributed quantum architectures plays a pivotal role for different applications. In particular, deterministic, robust, and fast protocols that prepare genuine multipartite entangled states are highly desirable. In this article, we propose a fractional quantum state transfer, in which the excitation of an emitter is partially transmitted through the quantum communication channel and then absorbed at a spatially separated node. This protocol is based on wavepacket shaping allowing for a fast deterministic generation of Bell states among two quantum registers and $W$ states for a general setting of $N$ qubits, either in a sequential or simultaneous fashion, depending on the topology of the network. By means of detailed numerical simulations, we show that genuine multipartite entangled states can be faithfully prepared within current experimental platforms and discuss the role of the main decoherence sources, qubit dephasing and relaxation, depending on the network topology. 
\end{abstract}

\maketitle

\section{Introduction}\label{s:intro}

Quantum networks offer a promising route towards scalable quantum-technological architectures~\cite{vanMeter,Wei22}. They consist of several spatially separated quantum registers (nodes) connected via quantum links or quantum communication channels (QCC) that support the exchange of quantum information on demand~\cite{Cirac99,vanMeter}. Quantum networks are of particular interest in various areas, ranging from quantum-information theoretical aspects~\cite{Acin07,Cavalcanti11,Perseguers10,Navascues20,ContrerasTejada21,Harney22} to applications such as the quest for a quantum internet~\cite{Kimble08,Wehner18}, distributed quantum computation~\cite{Cirac99,Caleffi22} or quantum memories~\cite{Giovannetti07}, among others.  As experimentally demonstrated, quantum nodes consisting of trapped ions or solid-state qubits can be linked to form quantum networks~\cite{Duan10,Northup2014,Reiserer15,Narla16,Kuzyk18,Axline18,CampagneIbarcq18,Leung19,Bienfait19,Nguyen19,Magnard20,Bhaskar20,Zhong21,Pompili21,Wu22,Storz23}, where either phononic or photonic quanta are carriers of quantum information.  In such architectures, genuine multipartite entangled states, i.e. states that are not separable under any bipartition~\cite{Horodecki09}, are key to obtain a quantum advantage in technological applications ~\cite{Hillery99,Bourennane01,Epping17,Toth14}. However, an efficient and scalable method to generate genuinely multipartite entanglement across a quantum network remains a challenging task.

Quantum state transfer protocols play a pivotal role~\cite{Cirac96} as they allow for the distribution of pre-existing entangled states across a network. Yet, generating multipartite entanglement among distant quantum registers necessitates other schemes~\cite{Wallnofer19,Wu22,MiguelRamiro23,Perseguers08,Perseguers13}. Moreover, fast and deterministic operations are highly desirable. In this context, single-traveling photons in superconducting quantum networks grant unique opportunities: The good controllability~\cite{Pechal14,Zeytinoglu15} is accompanied by small photon losses in transmission lines over large distances and fast protocol times, allowing for deterministic quantum operations~\cite{Kurpiers17,Besse18,Magnard20,Penas22,Qiu23, glaser2023}. Thus, these devices have emerged as major contestants in the quest for achieving large-scale quantum networks. 

In this work, we propose a scheme to generate genuine multipartite entangled states across a quantum network in a deterministic fashion employing an \textit{fractional quantum state transfer}.  This scheme is based on quantum communication via traveling single bosonic excitations, and, as we show, it is well suited to produce multipartite entanglement in the form of $N$-qubit $W$ states among an arbitrary number of quantum registers. Such registers can be separated in a broad range of distances, and therefore, the resulting entangled $W$ state can be relevant for distributed quantum computation, communication and metrological tasks~\cite{Joo02,Joo03,Agrawal06,WangJian07,Grafe14,Ng14,Zhu15}. The protocol provides fast operation times and can be adapted depending on the topology and connectivity of the quantum network, operating in either a sequential or a simultaneous fashion. We illustrate our scheme by detailed numerical simulations and show its robustness to the most pressing decoherence sources, namely, dephasing and qubit decay. The proposed protocols can be readily implemented in current experimental setups, such as in superconducting qubit networks~\cite{Kurpiers17,Magnard20,Storz23}.

The article is organized as follows. In Sec.~\ref{s:intro} we introduce the fractional quantum state transfer protocol, and particularize its implementation using wavepacket shaping for two experimentally relevant nodes, namely, i) a qubit and transfer resonator with a time-dependent coupling, and ii) a qubit with a controllable effective decay rate. In Sec.~\ref{s:GME} we show how to prepare a genuine multipartite entangled $W$ state in a deterministic fashion making use of fractional state transfers for a linear quantum network (Sec.~\ref{ss:lin}), while a star-like network is analyzed in Sec.~\ref{ss:star}, where we detail how to obtain a faster entanglement generation with modified controls. In Sec.~\ref{s:exp} we present numerical results supporting the good performance and suitability of the derived protocols to faithfully prepare multipartite entangled $W$ states taking into account the main relevant decoherence sources in these setups. Finally, the main conclusions of the work are summarized in Sec.~\ref{s:conc}.

\begin{figure}
    \centering
    \includegraphics[width=\linewidth]{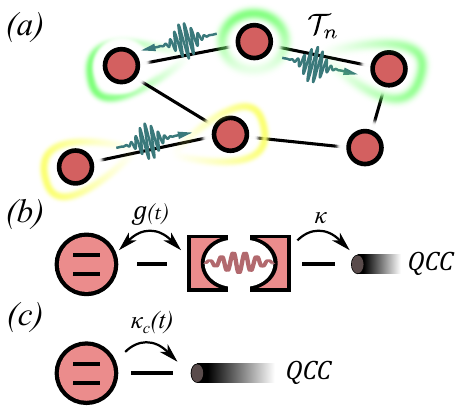}
    \caption{\small{(a) Schematic illustration of a quantum network. Different nodes (circles) are linked via quantum communication channels (QCCs) (solid lines). Fractional quantum state transfer $\mathcal{T}_n$ enables the exchange of a $1/n$ part of an excitation  between qubits placed within the nodes by performing emission and absorption protocols of the traveling excitation through the QCC, thus allowing for the distribution of multipartite entanglement across the network. Each node may consist of a (b) qubit-transfer resonator or a (c) qubit directly coupled to the QCC. In (b), the qubit-transfer resonator system, which decays into the QCC at rate $\kappa$, is coupled via a Jaynes-Cummings interaction with a controllable coupling $g(t)$, while in (c) the effective description of the decaying qubit allows for a controllable decay rate $\kappa_c(t)$. }}    
    \label{fig1}
\end{figure}

\section{Fractional quantum state transfer}\label{s:iqst}

We begin our analysis by introducing the fractional quantum state transfer protocol among two spatially separated qubits, which generalizes the seminal quantum state transfer process proposed in~\cite{Cirac96}. Initially, one qubit is prepared in a general state $\ket{\psi}=\alpha\ket{0}+\beta\ket{1}$, with $|\alpha|^2+|\beta|^2=1$. The fractional transfer operation $\mathcal{T}_n$ can be split into two stages: First, the emission of a fraction $1/n$ of the qubit excitation $\ket{1}$, with $n\geq 1$, mapping it onto a flying qubit through the QCC that travels until it reaches the second qubit. Second, $\mathcal{T}_n$ demands that the receiver qubit, initially in the state $\ket{0}$, absorbs the incoming excitation. In this manner, the operation $\mathcal{T}_n$ can be written as (up to local phases)
\begin{align}\label{eq:Tn}
\left(\alpha\ket{0}+\beta\ket{1}\right)\otimes \ket{0} \xrightarrow{\mathcal{T}_n} \alpha \ket{00}+\beta\sqrt{\frac{n-1}{n}}\ket{10}+\frac{\beta}{\sqrt{n}}\ket{01}.
  \end{align}
A standard quantum state transfer protocol is recovered for $n=1$, when the excitation of the first qubit is completely exhausted (see for example Refs.~\cite{Kurpiers17,Magnard20,Penas22}) and the state $\alpha\ket{0}+\beta\ket{1}$ is transferred to the second qubit.  However, contrary to the original quantum state transfer proposal~\cite{Cirac96}, the absorption protocol for $\mathcal{T}_n$ will not be the delayed time-reversed emission protocol for $n>1$. Importantly, in contrast to standard quantum state transfer, $\mathcal{T}_{n>1}$ directly enables a deterministic preparation of entangled states between the two connected qubits~\cite{CampagneIbarcq18,Axline18,Bienfait19,Leung19,Zhong21}. In particular, one obtains a maximal entangled state when $\alpha=0$ and $n=2$ in Eq.~\eqref{eq:Tn}. For $n\rightarrow\infty$, no excitation is transferred, and the initial state remains unchanged. The protocol $\mathcal{T}_n$ can be realized employing wavepacket shaping techniques in different architectures, as we detail in the following. To derive the protocols, it is assumed that the underlying Hamiltonian $\hat{H}$ of the full system (nodes and QCCs) conserves the total number of excitations. This is indeed a good description in most parameter regimes of state-of-the-art  realizations of quantum networks (see for example~\cite{Kurpiers17,Bienfait19,Magnard20,Storz23,Grebel24}). In Sec.~\ref{s:GME}, we will extend the scheme to multipartite systems, with or without simultaneous communication among nodes, and show how to generate multipartite entanglement across a quantum network (cf. Fig.~\ref{fig1}(a)).

\subsection{Time-dependent coupling between qubit and transfer resonator}\label{ss:qres}

Let us consider that the emitter node consists of a qubit coupled to a resonator, which in turn decays into the QCC at a rate $\kappa$ (see Fig.~\ref{fig1}(b)). Such a simple system constitutes the building block of existing quantum networks~\cite{Cirac96,Kurpiers17,Magnard20}. The elements in the node are considered to be resonant and coupled via a Jaynes-Cummings interaction with a time-dependent coupling $g(t)$~\cite{Zeytinoglu15}, assumed real, $g(t)\in \mathbb{R}$. The dynamics in the single-excitation subspace and in a rotating frame can be described within the Markovian approximation by the following equations of motion ($\hbar=1$)
\begin{align}\label{eq:qdot}
  \dot{q}(t)&=-ig(t)r(t),\\ \label{eq:rdot}
  \dot{r}(t)&=-ig^*(t)q(t)-\kappa r(t)/2.
\end{align} 
The complex parameters $q(t)$ and $r(t)$ denote the qubit and resonator single-excitation amplitudes, respectively. We consider that the qubit contains initially the excitation, $\lim_{t\rightarrow -\infty}|q(t)|^2=1$. The goal consists in injecting an excitation into the QCC, characterized by a shape $\gamma(t)$ such that $\int_{-\infty}^\infty |\gamma(t)|^2=n^{-1}$ so that $\lim_{t\rightarrow \infty}|q(t)|^2=(n-1)/n$.  Without loss of generality, the information is transmitted by a photon in a symmetric wavepacket with maximal bandwidth $\kappa$~\cite{Magnard20,Kurpiers17} but fulfilling the previous normalization, $\gamma(t)=\sqrt{\kappa/(4n)}{\rm sech}(\kappa t/2)$. Note that the specific choice of a sech-like wavepacket is not required (see~\cite{SM} for other photon shapes and narrower bandwidths), but it is chosen due to its popularity in previous works. From input-output theory, one identifies $|\gamma(t)|=\sqrt{\kappa}|r(t)|$. This allows us to reverse the equations of motion~\eqref{eq:qdot}-\eqref{eq:rdot} and obtain the control~\cite{Penas22,Penas23}, which now depends on $n$ and reads as~\cite{SM}
\begin{align}\label{eq:gt}
g(t;n)=\frac{\kappa{\rm sech}(\kappa t/2)}{2\sqrt{(n-1)(1+e^{\kappa t})^2+1}}.
\end{align}
Employing the control $g(t;n)$, the initial state $\ket{\psi(t\rightarrow-\infty)}=\ket{1}_q\ket{0}_r\ket{0}_{qcc}$ is transformed into $\ket{\psi(t\rightarrow\infty)}=\sqrt{\frac{n-1}{n}}\ket{1}_q\ket{0}_r\ket{0}_{qcc}+\frac{e^{i\phi}}{\sqrt{n}}\ket{0}_q\ket{0}_r\ket{1}_{qcc}$, where $\phi$ is a dynamical phase that can be controlled by a suitable design of the coupling $g(t;n)$. Here, $\ket{1}_q$, $\ket{1}_r$, and $\ket{1}_{qcc}$ denote a single excitation state in the qubit, resonator, and QCC, respectively. The traveling photon through the QCC can be then absorbed in the other end by another qubit-resonator system, but employing a time-reversed protocol setting $n=1$ with a delay due to the wavepacket propagation time $t_{d}=L/v_g$ through the QCC with length $L$ and group velocity $v_g$, i.e.  $g(-t+t_d;1)$. For simplicity, we assume an equal decay rate $\kappa$ and resonant frequency at the receiver end (see~\cite{SM} for the details when decay rates and frequencies differ).  Note that the absorption protocol is performed with $n=1$ to map the traveling photon into the qubit excitation at the reception end completely. In this manner, the emission  $g(t;n)$ and reception $g(-t+t_d;n=1)$ will complete the protocol $\mathcal{T}_n$ (cf. Eq.~\eqref{eq:Tn}). This generalization for $n\geq 1$ reduces to the standard control pulse $g(t)=\kappa/2 \ {\rm sech}(\kappa t/2)$ when setting $n=1$ in Eq.~\eqref{eq:gt} as used in the literature~\cite{Magnard20,Kurpiers17,Penas22}. 

In addition, it is worth mentioning that including a Purcell filter between the transfer resonator and the QCC non-trivially modifies the control $g(t;n)$. We refer the reader to~\cite{SM} for the details of the derivation and the time-dependent control in this situation.

\subsection{Time-dependent qubit decay rate}\label{ss:qdecay}
The emission of an excitation fraction, and thus the protocol $\mathcal{T}_n$, can also be achieved  by controlling the effective decay rate of the qubit directly into the QCC (see Fig.~\ref{fig1}(d))~\cite{Bienfait19,Grebel24}. In this particular case, the nodes in the network can be effectively described by a qubit with a decay rate $\kappa_c(t)$ so that the dynamics in the rotating frame reads as
\begin{align}
    \dot{q}(t)=-\kappa_c(t)q(t)/2.
\end{align}
Again, employing $\gamma(t)=\sqrt{\kappa/(4n)}{\rm sech}(\kappa t/2)$ and the input-output theory relation $|\gamma(t)|=\sqrt{\kappa_c(t)}|q(t)|$, one finds $\kappa_c(t)=|\gamma(t)|^2/(1-\int_{-\infty}^t dt'|\gamma(t')|^2)$ assuming $\lim_{t\rightarrow -\infty}|q(t)|^2=1$. This expression reduces to
\begin{align}
    \kappa_c(t;n)=\frac{\kappa {\rm sech}^2(\kappa t/2)}{2(2n-1-\tanh(\kappa t/2))}.
\end{align}
Note that $\kappa$ refers to the frequency bandwidth of the produced wavepacket, while $\kappa_c(t)$ is the controllable qubit decay rate. 
For $n>1$, $\kappa_c(t)$ features a maximum at $t\gtrsim 0$, whose value decreases with $n$. 
Following the same discussion as for the qubit coupled to a transfer resonator, in order to realize $\mathcal{T}_n$, the emission protocol is performed with $\kappa_c(t;n)$, while the receiver qubit requires a time-reversed protocol after a delay $t_d$ but with $n=1$, regardless of the emitted fraction, i.e. $\kappa_c(-t+t_d;n=1)$.

\section{Genuine multipartite entanglement distribution in quantum networks}\label{s:GME}

The fractional quantum state transfer $\mathcal{T}_n$ introduced in Sec.~\ref{s:iqst} can be generalized to the creation of multipartite entangled states in quantum networks comprising $N>2$ nodes. 
In particular, we focus on the genuine multipartite entangled $W$ state, which for $N$ qubits can be written as
\begin{align}\label{eq:Wn}
    \ket{W_N}=\frac{1}{\sqrt{N}}\sum_{j=1}^N|1_j\rangle,
\end{align}
where $|1_j\rangle$ denotes a single excitation in the $j$th qubit, and $\ket{0}$ in the rest, i.e. $|1_j\rangle=\hat{\sigma}_j^+\ket{{\rm vac}}$, where $\ket{{\rm vac}}$ is the vacuum state for all the system, and $\hat{\sigma}^+=\ket{1}\bra{0}$ the raising operator. 

For $N=2$, $\ket{W_2}$ simply reduces to a standard Bell state, while for $N=3$, $\ket{W_3}$ represents one of the two different families of genuinely tripartite entangled states~\cite{Duer00,Acin01,Verstraete02}, being the Greenberger-Horne-Zeilinger (GHZ) the other class. Although GHZ states are maximally entangled~\cite{Horodecki09}, $W$ states are more robust to decoherence~\cite{Chaves10} at the expense of displaying less entanglement. As mentioned above, $W$ states are useful in diverse applications, such as random number generation~\cite{Grafe14}, quantum key distribution~\cite{Zhu15,Joo02,WangJian07}, quantum teleportation~\cite{Joo03,Agrawal06} or even for quantum metrological purposes~\cite{Ng14}, where GHZ states appear as the canonical example to achieve an enhanced precision. 

\begin{figure}
    \centering
    \includegraphics[width=0.6\linewidth]{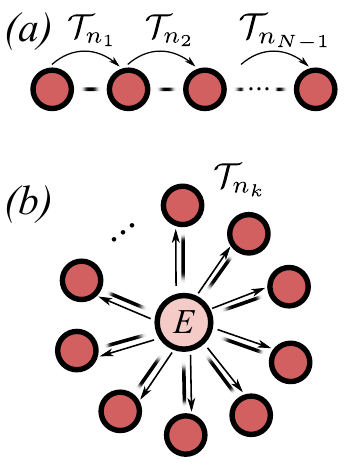}
    \caption{(a) Linear quantum network, where all nodes all connected via a unique path. In this case, a sequential application of fractional quantum state transfer $\mathcal{T}_{n_j}$ allows for the distribution of genuine multipartite entanglement. (b) Star quantum network, where a central node (emitter) is connected to all neighboring nodes via independent QCCs. This network connectivity allows for the simultaneous application of fractional quantum state transfers, and thus it accelerates the generation of genuine multipartite entangled states across the network. See main text for details. }
    \label{fig2}
\end{figure}

In the following, we detail how the state $\ket{W_N}$ can be obtained in two distinct quantum network topologies, namely, linear and star networks (cf. Fig.~\ref{fig2}). Depending on the connectivity, multipartite entanglement can be distributed either by a sequential or simultaneous protocol.


\subsection{Linear quantum network: Sequential protocol}\label{ss:lin}
In a linear quantum network, all the $N$ nodes are connected via a unique path, as depicted in Fig.~\ref{fig2}(a). For the communication to be propagated along the network, each of the nodes must be able to absorb excitations coming from its right and left neighbors. In particular, we assume that this interaction follows a similar scheme as before, with independent controls to either direction. That is, for nodes made of a qubit interacting via controllable coupling $g(t)$ with transfer resonators (cf. Fig.~\ref{fig1}(b)), there is a transfer resonator at either end, and the coupling between them and the qubit is defined as $g_{j,L}(t)$ and $g_{j,R}(t)$, where the subscript $j$ denotes the $j$th qubit in the network. Similarly, for nodes where wavepacket shaping is done via a controllable qubit decay rate, we distinguish $\kappa_{cj,L}(t)$ and $\kappa_{cj,R}(t)$, depending on the QCC side.

The linear quantum network is initially prepared (at time $t_0$) in a state containing a single excitation in the qubit at one of the ends, namely, $\ket{\psi(t_0)}=\ket{1_1}$. Ideally, for perfect emission and absorption protocols across the network, one can write the state at time $t$ as
\begin{align}
    \ket{\psi(t)}=\sum_{j=1}^Nq_j(t) |1_j\rangle,
\end{align}
since the number of excitations is conserved, and with a normalization $\sum_j|q_j(t)|^2=1$. Note that $q_1(t_0)=1$ and $q_{j\neq 1}(t_0)=0$. 

Now, for qubit-transfer resonator nodes, one chooses $g_{1,R}(t)=g(t;n_1)$ and $g_{2,L}(-t+t_{d,1};1)$ with $t_{d,1}=L_1/v_{g,1}$ the propagation time of the wavepacket through the first QCC of length $L_1$ and group velocity $v_{g,1}$. For a controllable qubit decay rate, one employs $\kappa_{c1,R}(t)=\kappa_c(t;n_1)$ and $\kappa_{c2,L}(t)=\kappa_c(-t+t_{d,1};1)$. Upon this first emission-absorption protocol that implements $\mathcal{T}_{n_1}$ among qubits 1 and 2, of duration $\tau_1$ one finds
the state $\ket{\psi(\tau_1+t_0)}$ with coefficients
\begin{align}
q_1(\tau_1+t_0)&=\sqrt{\frac{n_1-1}{n_1}},\\q_2(\tau_1+t_0)&=\frac{e^{i\phi_2}}{\sqrt{n_1}},\\q_{j>2}(\tau_1+t_0)&=0.
  \end{align}
Here $\phi_2$ denotes the relative phase accumulated during the evolution. 
  
Repeating the previous operation sequentially between neighboring nodes, i.e. performing  $\mathcal{T}_{n_2}$ between node 2 and 3, and so on (cf. Fig.~\ref{fig2}(a)), with $T=\sum_{j=1}^{N-1}\tau_j+t_0$ the final time upon the protocol, one arrives at a state $\ket{\psi(T)}$ with coefficients
\begin{align}
q_{j<N}(T)&=e^{i\phi_j}\sqrt{n_j-1}\prod_{k=1}^j \frac{1}{\sqrt{n_k}}, \\ q_N(T)&=e^{i\phi_N}\prod_{k=1}^{N-1}\frac{1}{\sqrt{n_k}}.
  \end{align}
  The previous expression can be easily manipulated to find a balanced superposition $|q_j|=1/\sqrt{N}$ for $j=1,\ldots,N$. This is achieved performing a sequential $N-1$ fractional quantum state transfers, namely, $\mathcal{T}_{n_1},\ldots,\mathcal{T}_{n_j},\ldots \mathcal{T}_{n_{N-1}}$ with $n_{k}=(N+1-k)/(N-k)$ and $k=1,\ldots,N-1$. Note that, by construction, $n_{k+1}>n_k>1$ with $n_{N-1}=2$,  meaning that as we move along the quantum network, the fraction of the transferred excitation gets smaller. This designed protocol leaves the quantum network deterministically in a $N$-qubit $W$ state, $\ket{W_N}$ (cf. Eq.~\eqref{eq:Wn}). Since it  requires $N-1$ fractional quantum state transfers, $2(N-1)$ processes of emission and absorption are needed.

  Distinct choices of $n_k$, initial states and additional gate capabilities in the network can lead to other interesting multipartite entangled states. For example,  $N/2$-qubit $W$ state for qubits placed at even sites, $\ket{\psi(T)}=\sqrt{2/N}\sum_{j=1}^{N/2}e^{i\phi_{2j}}|1_{2j}\rangle$, follows by simply alternating the transferred excitation fractions $n_k$ at even and odd sites.  In addition, one can immediately verify that for $n_1=2$ and $n_{j>1}=1$, one can prepare a Bell state among the first and last qubit in the linear quantum network, $|q_{1,N}|=1/\sqrt{2}$ and $q_{j\neq 1,N}=0$. This consists in transferring half excitation of the first qubit through the intermediate ones up to the last qubit, so that the final state reads $\ket{\psi(T)}=\left(\ket{10\ldots 0}+e^{i\phi_N}\ket{0\ldots 01}\right)/\sqrt{2}$. 
  Finally, let us mention that the accumulated phases $\phi_j$ can be accounted for by a suitable choice of the controls,  either by $g_j(t;n)\rightarrow g_j(t;n)e^{-i\phi_j}$ or by a frequency modulation of the qubits.

\subsection{Star quantum network: Simultaneous protocol}\label{ss:star}

The previous protocol is based on an sequential application of fractional quantum state transfers among neighboring nodes. Although a sequential protocol suffices to distribute entanglement, other network architectures may grant a larger connectivity, which can be harnessed to accelerate entanglement distribution across the network. As an example, we consider a star quantum network, as depicted in Fig.~\ref{fig2}(b), which consists of a central node linked to $N$ different nodes via $N$ QCCs. In this situation, we require that the central node, denoted here as emitter (E), is able to address each of the $N$ QCCs individually. Furthermore, it is assumed that distinct wavepackets can be injected through these QCCs simultaneously. In this manner, the controls derived in Sec.~\ref{s:iqst} are no longer suitable, as the emitter now transfers fractions of excitations to more than one node at the same time. This translates, therefore, in modifications of either $g_j(t;n)$ (for nodes made of qubit-transfer resonator) or $\kappa_{cj}(t)$ (for controllable qubit decay rate).

Let us first consider the scenario of nodes made of qubit plus transfer resonators. In this network configuration, the emitter consists of a qubit plus $N$ transfer resonators, each decaying with rate $\kappa_j$ into the $j$th QCC, while the receiver nodes are as depicted in Fig.~\ref{fig1}(b). The equations of motion in the rotating frame (assuming again resonant elements) read as
\begin{align}
    \dot{q}_E(t)&=-i\sum_{j=1}^Ng_j(t)r_j(t)\\
    \dot{r}_j(t)&=-ig_j^*(t)q_E(t)-\kappa_jr_j(t)/2,
\end{align}
where $q_E(t)$ and $r_j(t)$ denote the amplitude of having an excitation in the emitter and the $j$th resonator, respectively. Considering again a sech-like wavepacket, but now transferring a $1/n_j$ excitation through each of the $N$ QCCs, we have $\gamma_j(t)=\sqrt{\kappa_j/(4n_j)}{\rm sech}(\kappa_j t/2)$. Since $d/dt(|q(t)|^2+\sum_{j=1}^N|r_j(t)|^2)=-\sum_{j=1}^N\kappa_j |r_j(t)|^2$ it is possible to find~\cite{SM}
\begin{align}\label{eq:gsim}
g_j(t;{\bf n})=\frac{e^{\kappa_j t/2} \kappa_j}{(1+e^{\kappa_j t})^2\left[n_j(1-\sum_{k=1}^N\mathcal{K}_k(t)) \right]^{1/2} }
  \end{align}
with $\mathcal{K}_{k}(t)=[2+{\rm sech}^2(\kappa_k t/2)+2\tanh(\kappa_k t/2)]/(4 n_k)$, and ${\bf n}=(n_1,n_2,\ldots,n_N)$, such that $\sum_{k=1}^N n_k^{-1}\leq 1$. If no excitation is transferred through the $j$th QCC, $n_j\rightarrow \infty$, which implies  $\mathcal{K}_j(t)=0$ as well as $g_j(t;{\bf n})=0$, as expected. Moreover, if only an excitation fraction is transferred through the $j$th QCC ($n_{k\neq j}\rightarrow \infty$, so $\mathcal{K}_{k\neq j}(t)=0$), one falls back to the scenario discussed in Sec.~\ref{ss:qres}, and accordingly $g_j(t;{\bf n})$ in Eq.~\eqref{eq:gsim} simply reduces to Eq.~\eqref{eq:gt}.

For nodes where emission and absorption is done via a controllable qubit decay rate, we proceed in a similar manner. The emitter decays into $N$ QCCs with decay rate $\kappa_{c,j}(t)$,  which in the rotating frame corresponds to
\begin{align}
    \dot{q}_E(t)=-\sum_{j=1}^N\kappa_{c,j}(t)q_E(t)/2,
\end{align}
Employing the same sech-like wavepacket $\gamma_j(t)$ with bandwidth $\kappa_j$, and making use of the input-output relation $|\gamma_j(t)|=\sqrt{\kappa_{c,j}(t)}|q_(t)|$, one can manipulate the equations of motion to find a close expression for $\dot{\kappa}_{c,1}(t)$ in terms of $f_j(t)=\sqrt{\gamma_j(t)/\gamma_1(t)}$~\cite{SM}
\begin{align}
    \dot{\kappa}_{c,1}(t)=\left[\sum_{k=1}^N f_k^2(t)\right]\kappa_{c,1}^2(t)+2\kappa_{c,1}(t)\frac{\dot{\gamma}_1(t)}{\gamma_1(t)},
\end{align}
while $\kappa_{c,j}(t)=f_j^2(t)\kappa_{c,1}(t)$. For simplicity, we just provide here the expression when all bandwidths are equal, $\kappa_j=\kappa \ \forall j$, which becomes
\begin{align}\label{eq:kappasim}
    \kappa_{c,j}(t)=\frac{\kappa {\rm sech}^2(\kappa t/2)}{2n_j(2-(1-\tanh(\kappa t/2))\sum_{k=1}^N n_k^{-1})},
\end{align}
while the general case can be found in~\cite{SM}. 
For $n_j\rightarrow \infty$, $\kappa_{c,j}(t)=0$ and no excitation is injected into the $j$th QCC. Again, if an excitation fraction is emitted only through a unique QCC, then  Eq.~\eqref{eq:kappasim} reduces to Eq.~\eqref{eq:kappat} (cf. Sec.~\ref{ss:qdecay}). 

 As anticipated, the simultaneous emission through independent QCCs leads to a non-trivial modification of the controls, given in Eqs.~\eqref{eq:gsim} and~\eqref{eq:kappasim}. Although they have been derived considering strictly simultaneous emission of wavepackets that propagate through the QCCs, it is certainly possible to allow for arbitrary time delays among them. We refer, however, to~\cite{SM} for a discussion on this topic.

The simultaneous emission from the emitter towards the $N$ receiver nodes accelerates the protocol to distribute genuine multipartite entanglement. Again, the absorption in the receiver nodes must be done according to the time-reversed controls as detailed in Sec.~\ref{s:iqst}, namely, $g(-t+t_{d,j};n=1)$ and $\kappa_c(-t+t_{d,j};n=1)$ where $t_{d,j}$ refers to the propagation time through the $j$th QCC. The total time of the protocol is only limited by a single emission-absorption for the slowest QCC, and thus faster than the sequential scheme at the expense of requiring the synchronization among all nodes and a careful simultaneous control on the emitter. In particular, an initial state with $q_E(t_0)=1$ 
and $q_{j\neq E}(t_0)=0$, is mapped to $q_E(T)=(1-\sum_{j=1}^N n_j^{-1})^{1/2}$ and $q_{j}(T)=e^{i\phi_j}/\sqrt{n_j}$.  The emission with $\sum_kn_k^{-1}=1$ completely exhausts the emitter excitation, $q_E(T)=0$. Moreover, setting $n_j=N \ \forall j\in \{1,N\}$, i.e. for each of the $N$ emitted wavepackets, the final state upon absorption  corresponds to a $\ket{W_N}$ among the $N$ receiver nodes. For $n_j=N+1 \ \forall j\in\{1,N\}$, the emitter also participates in the entangled state, leading to a $\ket{W_{N+1}}$ state.

In the following, we focus on a particular setup to test the performance of the sequential and simultaneous protocols to prepare a genuine multipartite entangled state $\ket{W_3}$ for three qubits taking into account the dynamics of realistic QCCs. In addition, we detail the impact of the main pressing decoherence effects in these systems, namely, qubit decay and dephasing.

\section{Decoherence effects and feasibility}\label{s:exp}

\subsection{Model}\label{ss:model}

In order to illustrate the performance of the general framework put forward in Sec.~\ref{s:iqst} and~\ref{s:GME}, we consider two distinct quantum networks made of qubits coupled to transfer resonators operating in the microwave regime, close to the experimental setups reported in~\cite{Kurpiers17,Magnard20,Storz23}. In this case, the QCCs are embodied by waveguides at cryogenic temperatures that allow for the exchange of itinerant single microwave photons with minimal losses. 

We first consider a linear quantum network comprising three nodes, similar to the illustration in Fig.~\ref{fig2}(a). The Hamiltonian $H_l$, where the subscript $l$ stands for linear quantum network, can be written as
\begin{align}
    H_{l}=\sum_{i=1}^3H_{n,i}+\sum_{i=1,2}H_{qcc,i}+\sum_{i=1,2}H_{n-qcc,i},
\end{align}
being $H_{n,i}$ the local Hamiltonian for each node, $H_{qcc,i}$ the Hamiltonian for each QCC and $H_{n-qcc,i}$ the interaction between nodes and QCCs. 
The Hamiltonian for the nodes 1 and 3, i.e. $H_{n,1}$ and $H_{n,3}$, read as
\begin{align}\label{eq:Hni}
    H_{n,i}&=\delta_i \sigma_i^+\sigma_i^-+\omega_i a_i^\dagger a_i+g_i(t)(\sigma_i^+a_i+{\rm H.c.}) \quad i=1,3
\end{align}
where $\sigma^+$ and $\sigma^-$ are the usual raising and lowering spin operators, while  $a^\dagger$ and $a$ the bosonic creation and annihilation operators, such that $[a,a^\dagger]=1$, and $g_i(t)$ is the tunable coupling between the $i$th qubit and its transfer resonator. Recall that the couplings are assumed to be real. The central node containing qubit 2, is coupled to a left and right transfer resonators, namely,
\begin{align}
    H_{n,2}=\delta_2 \sigma_2^+\sigma_2^-+\sum_{s=L,R}\left(\omega_{2,s} a_{2,s}^\dagger a_{2,s}+g_{2,s}(t)(\sigma_2^+a_{2,s}+{\rm H.c})\right).
\end{align}
The QCCs connecting  qubit 1 with 2 and qubit 2 with 3, are represented by a collection of $N$ bosonic modes,
\begin{align}\label{eq:Hqcci}
    H_{qcc,i}=\sum_{k}\Omega_{k}b_{k,i}^\dagger b_{k,i},
\end{align}
fulfilling also $[b_k,b_{k'}^\dagger]=\delta_{k,k'}$, 
while the interaction is given by
\begin{align}
    H_{n-qcc,1}&=\sum_{k}G_k\left(b_{k,1}^\dagger a_1 +(-1)^k b_{k,1}^\dagger a_{2,L}+{\rm H.c.}\right)\\
    H_{n-qcc,2}&=\sum_{k}G_k\left(b_{k,2}^\dagger a_{2,R} +(-1)^k b_{k,2}^\dagger a_{3}+{\rm H.c.}\right).
\end{align}
The alternating sign in the coupling of the waveguide mode $k$ stems from the parity of the electromagnetic mode at its ends~\cite{Pozar}. Note that both QCCs are equivalent, i.e. they feature the same dispersion relation $\Omega_k$ and couplings $G_k$. In particular, following the setup employed in~\cite{Kurpiers17,Magnard20,Storz23}, we assume a WR90 waveguide with a non-linear dispersion relation $\Omega_k=c\sqrt{(\pi/l_c)^2+k^2}$, being $c$ the speed of light in vacuum and $l_c$ the cross-section length of the waveguide, $l_c=0.02286$ m. For simplicity, we assume resonant elements and equal decay rates of the resonators, $\delta_i=\omega_i=\omega$, $\kappa=\kappa_i$ for $i=1,2$ and $3$. In this manner, the couplings read $G_k=\sqrt{\kappa v_g \Omega_k/(2\omega L)}$ as customary in the Ohmic regime~\cite{ripoll2022}, where $v_g$ is the group velocity and $L$ is the length of the waveguide.

The second model we consider is a star-like quantum network, cf. Fig.~\ref{fig2}(b), consisting of four nodes, i.e. a central emitter plus three nodes. The Hamiltonian of this star network, $H_s$, can be written as
\begin{align}
    H_s=H_E+\sum_{i=1}^3\left(H_{n,i}+H_{qcc,i}+H_{n-qcc,i}\right).
\end{align}
The Hamiltonian of the emitter $H_{E}$ reads as
\begin{align}
    H_E=\delta_E\sigma_E^+\sigma_E^-+\sum_{i=1}^3 \omega_{E,i} a_{E,i}^\dagger a_{E,i}+\sum_{i=1}^3g_{E,i}(t)(\sigma_E^+a_{E,i}+{\rm H.c}),
\end{align}
while the receiver nodes are described by $H_{n,i}$ as given in Eq.~\eqref{eq:Hni} for $i=1,2$ and $3$, and similarly for $H_{qcc,i}$ (cf. Eq.~\eqref{eq:Hqcci}). The interaction between the QCCs, emitter and receiver nodes is given by
\begin{align}
    H_{n-qcc,i}=\sum_{k}G_k\left(b_{k,i}^\dagger a_{E,i} +(-1)^k b_{k,i}^\dagger a_{i}+{\rm H.c.}\right).
\end{align}
As in the previous case, we assume resonant elements so that the frequency of qubits and resonators is set to $\omega$, as well as an equal decay rate $\kappa$. The dispersion relation of the QCCs and the corresponding couplings is taken as for $H_l$ given above. 

Note that in both cases, the counter-rotating terms in the interaction Hamiltonians have been neglected by virtue of the rotating-wave approximation. This is a good approximation in this setup, as the frequencies ($\omega$ and $\Omega_k$) are typically two orders of magnitude larger than the coupling strength ($G_k$ and $g(t)$).

\subsection{Numerical simulations}\label{ss:num}

We benchmark the generation of a genuine multipartite entangled state $W$ employing a fractional quantum state transfer, and performing numerical simulations in the two quantum networks discussed above. In particular, we perform a sequential protocol for the linear quantum network, while the star network allows for the simultaneous control. In both cases, we aim at preparing a three-qubit $W$ state, i.e. $\ket{W_3}$. 

The numerical simulations take advantage of the conservation of the total excitation of $H_l$ and $H_s$. Considering an initial state containing a single excitation in the qubit 1 (linear) or in the emitter qubit (start), one can exactly simulate the coherent dynamics relying on the Wigner-Weisskopf Ansatz. In the case of the linear network, the wavefunction at all time reads as
\begin{align}
    \ket{\Psi_l(t)}=\bigg[ &\sum_{i=1}^3 q_i(t)\sigma^+_i+\sum_{i=1,3}c_i(t)a_i^\dagger  \nonumber \\&+\sum_{s=L,R}c_{2,s}(t)a_{2,s}^\dagger +\sum_{k,i=1,2}\psi_{k,i}(t)b_{k,i}^\dagger\bigg]\ket{{\rm vac}},
\end{align}
where the amplitudes fulfill the normalization $\sum_{i=1}^3|q_i(t)|^2+\sum_{i=1,3}|c_i(t)|^2+\sum_{s=L,R}|c_{2,s}(t)|^2+\sum_{k,i=1,2}|\psi_{k,i}(t)|^2=1$, and where $\ket{\rm vac}$ denotes the vacuum state for the whole system. Similarly, for the star quantum network, 
\begin{align}
\ket{\Psi_s(t)}=\bigg[ &q_E(t)\sigma_E^++\sum_{i=1}^3 q_i(t)\sigma^+_i+\sum_{i=1}^3(c_{E,i}(t)a_{E,i}^\dagger+c_i(t)a_i^\dagger) \nonumber\\&+\sum_{k}\sum_{i=1}^3\psi_{k,i}(t)b_{k,i}^\dagger\bigg]\ket{{\rm vac}}.
\end{align}
These Ans\"atze lead to a set of coupled differential equation in terms of the coefficients, e.g. $\frac{d}{dt}\ket{\Psi_l(t)}=-iH_l\ket{\Psi_l(t)}$, where the initial condition is $q_1(t_0)=1$ (linear) and $q_E(t_0)=1$ (star), and the rest of the coefficients set to zero. 

We incorporate the impact of decoherence effects, considering the most relevant sources in this setup, namely, qubit dephasing and relaxation. Photon loss may also be another important source of error. However, as we discuss later based on the recent experiments reported in Refs.~\cite{Storz23,Qiu23},  the error introduced by photon losses is estimated to be smaller than those of qubit dephasing and relaxation.  Due to the low-frequency noise affecting solid-state devices~\cite{Nakamura02,Paladino14}, we model qubit dephasing by a quasistatic fluctuation adding a random uncertainty in each of the qubit frequencies obeying a Gaussian distribution. For a single realization, the frequency of the $i$th qubit is shifted according to $\delta_i\rightarrow \delta_i+\delta\omega_i$ with
\begin{align}
    p(\delta \omega)=\frac{1}{\sqrt{2\pi}\sigma}e^{-\delta\omega^2/(2\sigma^2)}.
\end{align}
The dephasing time is related to the standard deviation of the Gaussian distribution as $\sigma=\sqrt{2}/T_2$. Assuming an equal $T_2$ for all the qubits results in an independent and identically distributed fluctuation $\delta\omega_i$ for all qubits in the network. 

The impact of qubit relaxation is estimated by integrating the population of each of the qubits during the time span of the protocol, $\tau_i=\int dt |q_i(t)|^2$, and consistently reducing the amplitude of the corresponding qubit. That is, the population of the $i$th qubit decreases by a factor $e^{-\tau_i/T_1}$, which is brought to the vacuum state. Again, we consider equal $T_1$ times for all the qubits across the network. However, we anticipate that dephasing and spontaneous decay will have a different impact depending on the network configuration, i.e. depending on a sequential or simultaneous protocol. 

Averaging over many stochastic realizations for the dephasing noise, one obtains an ensemble $\overline{\rho}_T$. Since we are interested in the resulting state for the three qubits participating in the entangled state, all the other elements in the system are traced out, i.e. $\overline{\rho}={\rm Tr}_r[\overline{\rho}_T]$, so that $\overline{\rho}$ is the reduced state for three qubits.  Note that, due to the nature of the process, only the states $\ket{100}$, $\ket{010}$, $\ket{001}$ and $\ket{000}$ will be relevant. 

We simulate numerically the dynamics of the protocol to generate a $\ket{W_3}$ in these two networks, considering $\omega=2\pi\times 8.407$ GHz for qubits and transfer resonators, $N=100$ modes for each of the waveguides centered around the central frequency $\omega$ and a total length of $L=5$ m. In addition, we introduce a correction on the qubit frequencies to account for the Lamb shift that stems from the interaction of the transfer resonators with the QCCs. The effective Lamb shift is derived from depletion experiments as reported in previous works~\cite{Penas22,Penas23,Penas24}, and find $\delta_{\rm LS}=0.0065\kappa$ for each of the nodes so that $\delta\rightarrow\delta+\delta_{\rm LS}$. In order to generate the $\ket{W_3}$ the time-dependent controls presented in Sec.~\ref{s:iqst} and Sec.~\ref{s:GME} are used. In particular, for the linear network $g_1(t)$ follows $g(t;n)$ given in Eq.~\ref{eq:gt} with $n=3/2$, $g_{2,L}(t)$ corresponds to $g(-t+t_d;1)$, while $g_{2,R}(t)$ again $g(t+t_s;n)$ with $n=2$ and finally $g_3(t)=g(-t+t_d+t_s;1)$. Here $t_s$ is the time separation between the absorption from qubit 1 into qubit 2, and the following reemission from qubit 2. For the star network, the emitter coupling $g_{E,i}(t)$ follows Eq.~\eqref{eq:gsim} with ${\bf n}=(3,3,3)$, while the receivers a set to $g(-t+t_d;1)$. We investigate the interplay of the decay rates, which also fixes the photon bandwidth, namely $\kappa=2\pi\times 10$ MHz and $2\pi\times 50$ MHz, with the coherence times $T_1$ and $T_2$ in the range of $\sim 1\mu$s to $100\mu$s.  The total duration of the sequential protocol is set to $35/\kappa$, while the simultaneous allows for a faster generation, $14/\kappa$.

\begin{figure}
    \centering
    \includegraphics[width=0.9\columnwidth]{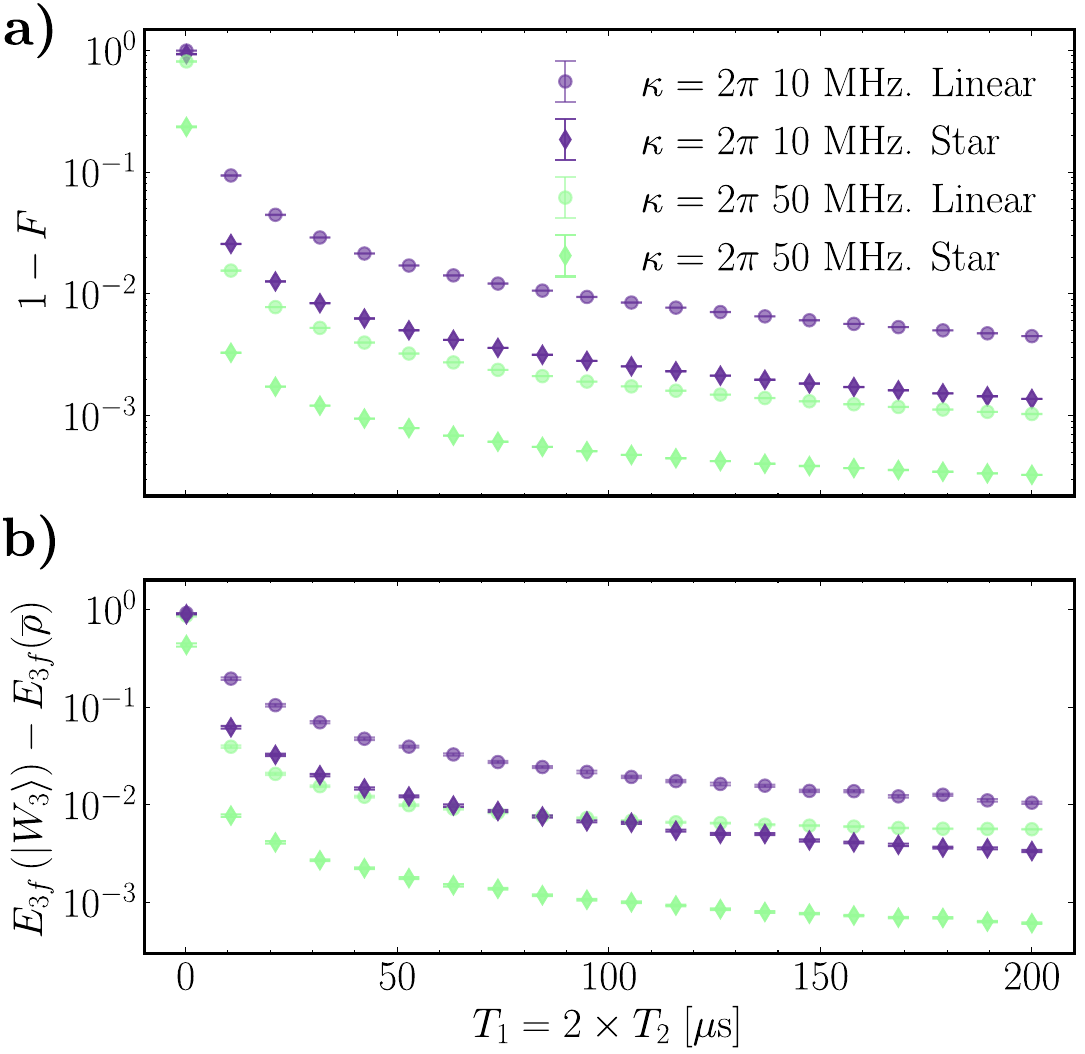}
    \caption{\small{Performance of the generated genuine multipartite entangled state $\ket{W_3}$. Panel (a) shows the infidelity $\log_{10}(1-F)$ with respect to the targeted state $\ket{W_3}$ for a sequential (linear) and simultaneous (star) protocols as a function of $T_2$ and $T_1$, setting $T_1=2T_2$, and for two different decay rates $\kappa$. Panel (b) shows the entanglement of formation $E_{3f}$ of the resulting state, in particular, the difference with the ideal value $E_{3f}(\ket{W_3})\approx 0.918$, i.e. $\log_{10}(E_{3f}(\ket{W_3})-E_{3f}(\overline{\rho}))$. In both panels, $\kappa =2\pi\times 50$ MHz (green color) provides better results as the protocol is faster than $\kappa=2\pi\times 10$ MHz (violet color). Each of the points corresponds to  $10^3$ stochastic realization to correctly include dephasing noise. Error bars correspond to the standard deviation for each case performing bootstrapping. See main text for details of the parameters. }}
    \label{fig3}    
\end{figure}

We quantify the performance of the protocol in terms of the fidelity $F$ with respect to the target $\ket{W_3}$ and entanglement of formation $E_{3f}$~\cite{Wootters98,Szalay15}. The fidelity $F$ is computed as $F=\max_{\phi_2,\phi_3}F(\phi_2,\phi_3)$ where $F(\phi_2,\phi_3)=\langle W_3'|\overline{\rho}|W_3'\rangle$ with $|W_3'\rangle=\frac{1}{\sqrt{3}}(\ket{100}+e^{i\phi_2}\ket{010}+e^{i\phi_3}\ket{001})$, i.e. allowing for local phases. The entanglement of formation $E_{3f}$ for a three-qubit pure state $\ket{\psi}$ is defined as $E_{3f}(\ket{\psi})=\min\{S_1,S_2,S_3 \}$ where $S_j$ is the von Neumann entropy for the reduced state of the $j$th qubit, i.e. $S_j=-{\rm Tr}[\rho_j \log_2\rho_j]$. For the $\ket{W_3}$, it follows $E_{3f}(\ket{W_3})=\log_2(3)-2/3\approx 0.918$. For a mixed state $\rho$, however, the entanglement of formation is computed as the minimum over all possible ensemble decomposition of $\rho$, i.e. $E_{3f}(\rho)=\min \sum_i p_i E_{3f}(\ket{\psi_i})$ where $\rho=\sum_i p_i \ket{\psi_i}\bra{\psi_i}$~\cite{Wootters98,Szalay15} (see~\cite{SM} for further details). 

In Fig.~\ref{fig3} we show the numerical results for the infidelity (a) and the entanglement of formation (b) as a function of coherence times $T_2=T_1/2$, and for the two networks with $\kappa=2\pi\times 10$ MHz and $2\pi\times 50$ MHz. The results have obtained averaging $1000$ stochastic realizations for the dephasing, while the error bars correspond to a standard deviation performing bootstrapping. As it can be seen, the simultaneous protocol performed in the star network is more robust to decoherence than its linear counterpart. This is expected since the star network allows for a faster protocol to generate the genuine multipartite entangled state $\ket{W_3}$. In particular, we find good results for state-of-the art coherence times, as evidenced by the entanglement of formation or the fidelity, $1-F<10^{-2}$ for $T_2>40\mu$s or even below $10^{-3}$ for the star network already for $T_2\gtrsim 20\mu$s. In addition, it is worth mentioning that a fidelity $F>2/3$ witnesses multipartite entanglement in the $W$ class~\cite{Walter16}. This is achieved for very short coherence times ($T_2,T_1\sim 5\mu$s). 

 Finally, we estimate the impact of photon loss through the QCCs taking into account the measurements made in two recent independent experiments. In~\cite{Storz23}, the measured photon loss was $0.5-1$ dB/Km, or $0.69\%$ loss for a $60$ m waveguide. On the other hand, a lower attenuation was reported in~\cite{Qiu23},  $0.32$ dB/Km, corresponding to a $0.4\%$ loss probability for its setup with a $60$ m long waveguide. These numbers show that photon loss in the considered example of a QCC with length $L=5$ m would provide an error $\sim 10^{-4}$. Therefore, photon loss would not hinder the fidelity of the process, even in the regime of a large $T_1$ and $T_2$ coherence times where the error is still larger than or comparable to $\sim 10^{-4}$.

\section{Conclusions and outlook}\label{s:conc}


In this work, we have proposed a fractional quantum state transfer protocol that allows for the generation of entangled states across a quantum network. The protocol builds on wavepacket shaping to inject a traveling single-photon through a quantum communication channel. The mechanism is first presented among two quantum nodes, allowing for the distribution of maximally-entangled states, considering two paradigmatic cases, i) a node comprises a qubit coupled to a transfer resonator with a tunable control, ii) a tunable time-dependent decay rate of a qubit into the quantum communication channel. The protocol is then extended to $N$ nodes, where a sequential application of fractional quantum state transfers can lead to a $W$ multipartite entangled state. In addition, we also consider a scenario where the control allows for a simultaneous emission, which speeds up the generation of $W$ states. The general framework put forward in the article is tested numerically, considering a linear and a star network configuration, and including the presence of the main sources of decoherence in these systems, namely, qubit dephasing and spontaneous decay. The numerical results show that multipartite entanglement can be faithfully realized in state-of-the-art setups, with overall fidelities in the range of $1-F\approx 10^{-2}$ or better.

This work opens up exciting future directions to develop deterministic and fast operations among distant nodes in quantum networks. Indeed, utilizing several photonic excitations together with multiplexed QCCs may entail novel opportunities for quantum computation and communication tasks.

\begin{acknowledgments}
 This work has been supported by the European Union's Horizon 2020 FET-Open project SuperQuLAN (899354), Proyecto Sin{\'e}rgico CAM 2020 Y2020/TCS-6545 (NanoQuCoCM), the Spanish project PID2021-127968NB-I00 (MCIU/AEI/FEDER,EU), and the CSIC Interdisciplinary Thematic Platform (PTI) Quantum Technologies (PTI-QTEP).
\end{acknowledgments}

\bibliography{paper.bib}


\widetext
\clearpage
\begin{center}
\textbf{\large Supplemental Material\\ Deterministic multipartite entanglement via fractional state transfer across quantum networks \vspace{0.2cm}}
\end{center}
\setcounter{equation}{0}
\setcounter{figure}{0}
\setcounter{table}{0}
\setcounter{page}{1}
\setcounter{section}{0}
\makeatletter
\renewcommand{\theequation}{S\arabic{equation}}
\renewcommand{\thefigure}{S\arabic{figure}}
\renewcommand{\bibnumfmt}[1]{[S#1]}
\renewcommand{\citenumfont}[1]{S#1}

\begin{center}
  G. F. Pe\~nas$^{1}$, J.-J. Garc\'ia-Ripoll$^{1}$, and R. Puebla$^{2}$\\
  {\em \small{$^{1}$Instituto de F{\'i}sica Fundamental (IFF), CSIC, Calle Serrano 113b, 28006 Madrid, Spain}}\\
  {\em \small{$^2$Departamento de F{\'i}sica, Universidad Carlos III de Madrid, Avda. de la Universidad 30, 28911 Legan{\'e}s, Spain}}
\end{center}

\section{Derivation of the control pulse}
\label{app:pulse}

\subsection{Time-dependent coupling between qubit and a transfer resonator}

We first discuss the where a qubit is coupled to a transfer resonator via a time-dependent control which leaks into the quantum communication channel at rate $\kappa$. This allows to map the qubit excitation into the QCC with a desired shape. The Hamiltonian in the interaction picture of the free energy terms, and considering a resonant qubit-resonator condition, is given by $H=g(t)\sigma^+a +{\rm H.c.}$, so that within the single-excitation subspace the wavefunction reads as $\ket{\psi(t)}=\left(q(t)\sigma^++r(t)a^\dagger\right)\ket{{\rm vac}}$ with $\ket{{\rm vac}}$ the vacuum state for the combined qubit-resonator system and $|q(t)|^2+|r(t)|^2=1\ \forall t$. The equations of motion for the amplitudes $q(t)$ and $r(t)$ are then given by
\begin{align}
  \dot{q}(t)&=-ig(t)r(t)\\
  \dot{r}(t)&=-ig^*(t)q(t)-\kappa r(t)/2
  \end{align}
so that the output photon is $|\gamma(t)|=\sqrt{\kappa}|r(t)|$. We aim at emitting a photon with a sech-like shape,
\begin{align}\label{eqSM:sechphot}
|\gamma(t;n)|=\frac{\sqrt{\kappa}}{\sqrt{4n}}{\rm sech}\left(\frac{\kappa t}{2}\right),
  \end{align}
with $n\geq 1$, so that a photon is emitted with probability $1/n$,
\begin{align}
\int_{-\infty}^{\infty} dt |\gamma(t;n)|^2 =\frac{1}{n}.
  \end{align}
From $d(|q(t)|^2+|r(t)|^2)/dt=-\kappa|r(t)|^2$, one finds $|q(t)|^2=|q(t_0)|^2-\frac{1}{\kappa}|\gamma(t;n)|^2-\int_{t_0}^td\tau |\gamma(\tau;n)|^2$, setting a  constraint on the photon shape since it must fulfill $ 0\leq |q(t)|^2\leq 1$. 
Setting $|q(t\rightarrow -\infty)|^2=1$, we arrive to the expression given in the main text,
\begin{align}\label{eqSM:gtn}
g(t;n)=\frac{e^{\kappa t/2}\kappa}{(1+e^{\kappa t})\sqrt{(n-1)(1+e^{\kappa t})^2+1}}.
\end{align}
Under the previous control $g(t;n)$ the qubit remains excited with a probability $|q(t\rightarrow\infty)|^2=(n-1)/n$. 
For $n=1$, we recover the standard form $g(t;1)=\kappa {\rm sech}(\kappa t/2)/2$, as expected, and the qubit excitation is totally exhausted. Note that for any $n>1$, the control is modified in a non-trivial fashion with respect to $g(t;1)$: For $\kappa t\gg 1$, $g(t;n>1)$ decays faster than $g(t;n=1)$, while the maximum scales as $\max_t g(t;n)\approx \kappa/\sqrt{9n}$ for $n\gg 1$ (see Fig.~\ref{figSM1}(a) for an illustration of $g(t;n)$). 

\subsubsection{Lorentzian-shaped photon}
The previous procedure can be applied to other photon shapes, as long as they fulfill $ 0\leq |q(t)|^2\leq 1$. As an example, we consider a Lorentzian shaped photon,
\begin{align}
|\gamma(t;n)|=\left(\frac{\kappa}{n\pi(1+\kappa^2 t^2)}\right)^{1/2},
  \end{align}
and as before, $\int_{-\infty}^{\infty} dt |\gamma(t;n)|^2 =n^{-1}$. This leads to
\begin{align}
  |g(t;n)|=\frac{\kappa(\kappa t-1)^2}{(1+\kappa^2 t^2)\sqrt{(4n-2)\pi(1+\kappa^2 t^2)-4(1+\kappa^2t^2){\rm atan}(\kappa t)-4}},
  \end{align}
which fulfills the physical condition $0\leq |q(t)|^2\leq 1$ for all $\kappa$ and $n\geq 1$. See Fig.~\ref{figSM1}(b) for an illustration of this derived pulse $g(t;n)$. 

\subsubsection{Gaussian-shaped photon}
\begin{figure}
    \centering
    \includegraphics[width=0.9\linewidth]{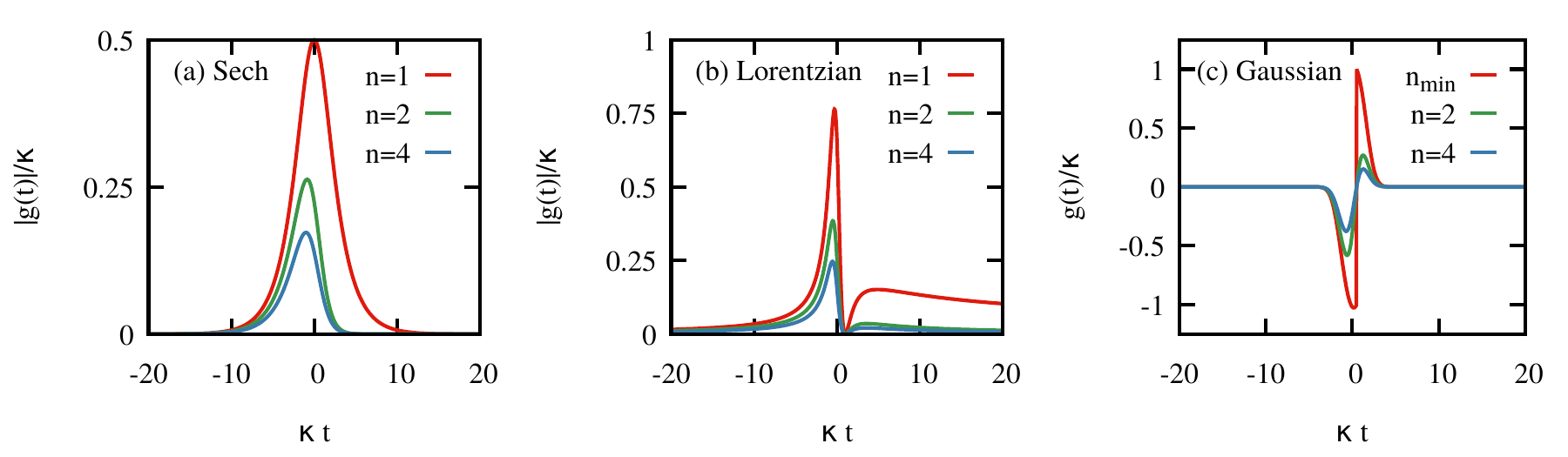}
    \caption{\small{Time-dependent control $g(t;n)$ for different targeted photon shapes, sech-like photon $|\gamma(t;n)|=\sqrt{\kappa/(4n)}{\rm sech}(\kappa t/2)$ (a), Lorentzian $|\gamma(t;n)=(\kappa/(n\pi(1+\kappa^2 t^2)))^{1/2}$ (b) and Gaussian $|\gamma(t;n)|=\sqrt{\kappa/\pi^{1/2}n}e^{-\kappa^2 t^2/2}$ (c), and for different values of the transferred fraction $n$. Note that a Gaussian photon can only be emitted with a probability smaller than $1/n_{\rm min}\approx 0.83$.}}
    \label{figSM1}
\end{figure}

Finally, we analyze a Gaussian-shaped photon,
\begin{align}
  |\gamma(t;n)|=\sqrt{\frac{\kappa}{\pi^{1/2}n}}e^{-\kappa^2 t^2/2}.
    \end{align}
Following the same steps we arrive to
\begin{align}
|q(t)|^2=\frac{1}{2n}\left(2n-1-\frac{2e^{-\kappa^2 t^2}}{\sqrt{\pi}}-{\rm erf}(\kappa t)\right)
  \end{align}
where ${\rm erf}(x)=\frac{2}{\pi^{1/2}}\int_0^xdt\ e^{-t^2}$. The previous equation is clearly non-physical for $n=1$ since $|q(t)|^2<0$ for some $t$. The minimum value of $|q(t)|^2$ takes place at $2\kappa t=1$, and therefore it is easy to find that the condition $0\leq |q(t)|^2\leq 1$ is fulfilled as long as  $n\geq n_{\rm min}$, with 
\begin{align}
  n_{\rm min}=\frac{1}{2}\left(1+\frac{2}{e^{1/4}\sqrt{\pi}}+{\rm erf}(1/2)\right)\approx 1.2.
\end{align}
For $n\geq n_{\rm min}$ the control pulse reads as
\begin{align}
g(t;n)=\frac{\kappa e^{-\kappa^2 t^2/2}(2\kappa t-1)}{\sqrt{2\pi^{1/2}(2n-1-{\rm erf}(\kappa t))-4e^{-\kappa^2 t^2}}}.
  \end{align}
Note that this control is negative in the range $t\in [-\infty,1/2\kappa)$ (see Fig.~\ref{figSM1}(c)). A Gaussian-shaped photon can be emitted with a probability smaller than $1/n_{\rm min}\approx 0.83$. Yet, the traveling excitation cannot be fully absorbed at the other end since it requires $n=1$, which makes Gaussian protocols not suitable for fractional quantum state transfer.

\subsubsection{Reduced bandwidth sech-shaped photons}
In certain circumstances, it may more convenient to inject photons with a narrower bandwidth. This can reduce potential distortion effects introduced by the propagation medium~\cite{Penas22}, or allow for asymmetric quantum networks where the decay rates among different nodes differ. For that, we introduce the following photon shape
\begin{align}
|\gamma(t,\eta;n)|=\sqrt{\frac{\kappa}{4\eta n}}{\rm sech}\left(\frac{\kappa t}{2\eta} \right),
  \end{align}
so that $\int_{-\infty}^{\infty}dt |\gamma(t,\eta;n)|^2=n^{-1}$, and whose bandwidth $\kappa/\eta$ is reduced by a factor $\eta>1$ with respect to the sech-photon given in the main text. This implies, however, a longer protocol duration. Following previous steps we find
\begin{align}\label{eq:gtem_eta}
g(t,\eta;n)=\frac{(\kappa/\eta n)^{3/2}n {\rm sech}(\kappa t/(2\eta))(\eta-\tanh(\kappa t/(2\eta)))}{2\sqrt{\frac{2\eta(2n-1-\tanh(\kappa t/(2\eta)))-{\rm sech}^2(\kappa t/(2\eta)}{\eta n}} }.
  \end{align}
Note that the control is not simply Eq.~\eqref{eqSM:gtn} with $\kappa\rightarrow \kappa/\eta$, since the rate at which the transfer resonator decays is still $\kappa$. 

This allows to investigate the situation where the transfer resonator receiving the photon with shape $|\gamma(t)|=\sqrt{\kappa/4}{\rm sech}(\kappa t/2)$ has a larger decay rate than the bandwidth of the photon, $\kappa_e\geq \kappa$. From previous results, one can find that then the control in the node that absorbs the excitation reads as~\cite{Penas22}
\begin{align}
g(t,\kappa,\eta;n=1)=\frac{\kappa {\rm sech}(\kappa t/2)(\eta-\tanh(\kappa t/2))}{4\sqrt{\frac{e^{\kappa t}(\eta-1)+\eta}{(1+e^{\kappa t})^2}}}.
  \end{align}
  where here $\eta=\kappa_e/\kappa\geq 1$. 
Note, however, that if $\kappa_e<\kappa$ it is not possible to perfectly absorb such wide photon with the limited bandwidth $\kappa_e$. This would require an emission employing Eq.~\eqref{eq:gtem_eta} to reduce the bandwidth of the traveling photon.

\subsection{Purcell filter between resonator and waveguide}
We now consider another scenario where a Purcell filter is placed between the resonator and the waveguide, as typically done in experimental setups to protect the qubit lifetime. Considering resonant elements, we find
\begin{align}
    \dot{q}(t)&=-i g(t) r(t)\\
   \dot{r}(t)&=-i g^*(t) q(t) -i g_p f(t)\\
    \dot{f}(t)&=-ig_p r(t)-\kappa f(t)/2
\end{align}
where now the coefficient $f(t)$ refers to the amplitude of one excitation in the Purcell filter, and $g_p$ a real coupling strength between resonator and filter, which in turn decays into the QCC at rate $\kappa$. Proceeding as before for a sech-like photon, $\gamma(t;n)=\sqrt{\kappa/(4n)}{\rm sech}(\kappa t/2)$, we can obtain the required $g(t)$. In particular, considering real parameters and $g_p>0$ for simplicity, we find 
\begin{align}
    r(t)=-\frac{1}{g_p}\left(\frac{\dot{\gamma}(t;n)}{\sqrt{\kappa}}+\frac{\sqrt{\kappa}}{2}\gamma(t;n)\right)
\end{align}
using $|\gamma(t;n)|=\sqrt{\kappa}|f(t)|$. Now, making also use of the relation $d/dt (|q(t)|^2+|r(t)|^2+|f(t)|^2)=-\kappa |f(t)|^2$, we arrive to 
\begin{align}
    |q(t)|^2&=|q(t_0)|^2-\frac{1}{g_p^2}\left(\frac{\dot{\gamma}(t;n)}{\sqrt{\kappa}}+ \frac{\sqrt{\kappa}}{2}\gamma(t;n)\right)^2-\frac{|\gamma(t;n)|^2}{\kappa}-\int_{t_0}^{t}d\tau \  |\gamma(\tau;n)|^2.
\end{align}
For the considered photon, the resonator amplitude takes the form (up to phases)
\begin{align}
    r(t)=-\frac{e^{\kappa t/2}\kappa }{(1+e^{\kappa t})^2g_p\sqrt{n}}.
\end{align}
Since $0\leq |r(t)|^2\leq 1$, and the maximum of $|r(t)|^2$ takes place at $t_m=-\log(3)/\kappa$, we find
\begin{align}
    |r_m|^2\equiv |r(t_m)|^2=\frac{3\kappa^2}{256 g_p^2 n}
\end{align}
so that 
\begin{align}\label{eq:km_cm}
    \kappa\leq \frac{16 g_p \sqrt{n}}{3}.
\end{align}
in order to ensure the physical constraint, which sets a relation between $\kappa$, $g_p$ and $n$. 

On the other hand, the qubit population becomes
\begin{align}
    |q(t)|^2=1-\frac{e^{\kappa t}((1+e^{\kappa t})^2(2+e^{\kappa t})g_p^2+\kappa^2}{(1+e^{\kappa t})^4 g_p^2 n}.
\end{align}
As before, this quantity has to be bounded between $0$ and $1$. For $n=1$, the setup must satisfy
\begin{align}\label{eq:kgpn1}
    \kappa\leq 2g_p.
\end{align}
For $n>1$, the physical constraint leads to a quite complicated expression relating $\kappa$ and $g_p$, yet, it can be shown that Eq.~\eqref{eq:kgpn1} is more restrictive than the one resulting for $n>1$ and the condition in Eq.~\eqref{eq:km_cm}. For example, for $n=2$, we find $\kappa\lesssim 3.8 g_p$. Finally, we can obtain the control from $g(t)=-\dot{q}(t)/f(t)$ (up to phase factors), so that
\begin{align}
g(t;n)=\frac{e^{\kappa t}(2(1+e^{\kappa t})^2g_p^2+(1-3e^{\kappa t})\kappa^2)\cosh(\kappa t/2)}{(1+e^{\kappa t})^4g_p\sqrt{n-\frac{e^{\kappa t}((1+e^{\kappa t})^2(2+e^{\kappa t})g_p^2+\kappa^2) }{(1+e^{\kappa t})^4 g_p^2}}}.
\end{align}
Again, the factor $n$ appears in a non-trivial manner in the control.

\subsection{Qubit directly coupled to the waveguide}
Finally, we consider the scenario where a qubit is directly coupled to the bosonic media through which an engineered bosonic excitation propagates. This was considered and experimentally realized in Ref.~\cite{Bienfait19}. The simple equation of motion for the qubit reads $\dot{q}(t)=-\kappa(t)q(t)/2$ where now photon shaping requires a time-dependent decaying rate $\kappa(t)$. Since the photon is given by $\gamma(t)=\sqrt{\kappa(t)}q(t)$, we obtain
\begin{align}\label{eq:kappat}
    \dot{\kappa}(t)=\kappa^2(t)+2\kappa(t)\frac{\dot{\gamma}(t;n)}{\gamma(t;n)}
\end{align}
Imposing $\gamma(t;n)=\sqrt{\kappa_p/(4n)}{\rm sech}(\kappa_p t/2)$, the time-dependent decay rate adopts the form 
\begin{align}\label{eq:kappaQWV}
    \kappa(t;n)=\frac{\kappa_p {\rm sech}^2(\kappa_p t/2)}{4n-2\tanh(\kappa_p t/2)-2}
\end{align}
upon imposing that $\lim_{t\rightarrow-\infty}|q(t)|^2=1$ and $\lim_{t\rightarrow \infty} |q(t)|^2=(n-1)/n$, and
where $\kappa_p$ refers to the frequency width of the injected photon. Exhausting the qubit excitation, $n=1$, we find $\kappa(t;n=1)=\kappa_p/2(1+\tanh(\kappa_p t/2))$, with $\kappa(t\rightarrow \infty;n=1)=\kappa_p$ its maximum value. For $n>1$, however, $\kappa(t\rightarrow \infty;n>1)=0$ with a maximum value $\kappa_m=\kappa_p(2n-1-2\sqrt{n^2-n})$. 


\section{Synchronous emission through more QCCs}
\subsection{Qubit-resonator nodes}
Considering a controllable qubit-resonator coupling per QCC between the qubit and another node, we can design controls that distribute an excitation across the network following similar steps as before. For $N$ QCCs the emitter is coupled to $N$ resonators that in turn decay in their corresponding QCC. The equations of motion are now given by (assuming real controls):
\begin{align}
    \dot{q}(t)&=-i\sum_{j=1}^Ng_j(t)r_j(t)\\
    \dot{r}_j(t)&=-ig_j^*(t)q(t)-\frac{\kappa_j}{2}r_j(t)
\end{align}
This allows to compute $|q(t)|^2$ from $d/dt(|q(t)|^2+\sum_{j=1}^N|r_j(t)|^2)=-\sum_{j=1}^N\kappa_j|r_j(t)|^2$ assuming $|\gamma_j(t;n_j)|=\sqrt{\kappa_j/(4n_j)}{\rm sech}(\kappa_j t/2)$. Once $|q(t)|^2$ is determined, the control $g_j(t)$ can be computed from the differential equation for $\dot{r}_j(t)$. After some calculations, the controls can be obtained as
\begin{align}
    g_j(t;{\bf n})=\frac{e^{\kappa_j t/2}\kappa_j}{(1+e^{\kappa_j t})^2\left[n_j(1-\sum_{k=1}^N(2+{\rm sech}^2(\kappa_k t/2)+2\tanh(\kappa_kt/2))/(4n_k) \right]^{1/2}}
\end{align}
which corresponds to the expression given in the main text, upon the definition $\mathcal{K}_k(t)=(2+{\rm sech}^2(\kappa_k t/2)+2\tanh(\kappa_kt/2))/(4n_k)$. 

For time-delays among wavepackets, $\gamma_j(t)=\sqrt{\kappa_j/(4n_j)}{\rm sech}(\kappa_j(t-t_j)/2)$, the modification is minimal: It just introduces a time-shift on the quantities $\mathcal{K}_k(t)=(2+{\rm sech}^2(\kappa_k (t-t_k)/2)+2\tanh(\kappa_k(t-t_k)/2))/(4n_k)$ so the general expression can be written as
\begin{align}
    g_j(t;{\bf n})=\frac{e^{\kappa_j (t-t_j)/2}\kappa_j}{(1+e^{\kappa_j t})^2[n_j(1-\sum_{k=1}^N\mathcal{K}_k(t))]^{1/2}}.
\end{align}

\subsection{Qubit directly coupled to waveguide}
As before, let us consider a qubit connected to two QQCs with controllable time-dependent decay rates $\kappa_1(t)$ and $\kappa_2(t)$. We aim at introducing distinct photons $\gamma_j(t;n_j)$ for $j=1,2$ so that $\int_{-\infty}^\infty dt |\gamma_j(t;n_j)|^2=n_j^{-1}$. The equation of motion in this case reads as
\begin{align}
    q(t)=-(\kappa_1(t)+\kappa_2(t))q(t)/2
\end{align}
with $\gamma_j(t)=\sqrt{\kappa_j(t)}q(t)$. Hence, we can write
\begin{align}
    q(t)=\frac{\gamma_1(t;n)+\gamma_2(t;n)}{\sqrt{\kappa_1(t)}+\sqrt{\kappa_2(t)}}
\end{align}
Now, we can write
\begin{align}
    \frac{\gamma_1(t;n_1)}{\gamma_2(t;n_2)}=\sqrt{\frac{\kappa_2(t)}{\kappa_1(t)}}=f(t) 
\end{align}
where $f(t)$ is a function that may depend on time, as well as on $\kappa_{p,j}$ and $n_j$. We can now equate
\begin{align}
    \dot{q}(t)=\frac{d}{dt}\left(\frac{\gamma_1(t;n_1)}{\sqrt{\kappa_1(t)}} \right)
\end{align}
with 
\begin{align}
\dot{q}(t)=-(\kappa_1(t)+\kappa_2(t))q(t)/2=-\kappa_1(t)(1+f^2(t)) q(t)/2=-\sqrt{\kappa_1(t)}(1+f^2(t))\gamma_1(t;n_1)/2.
\end{align}
This results in a differential equation for $\kappa_1(t)$ in terms of $f(t)$ and $\gamma_1(t;n_1)$, namely,
\begin{align}
    \dot{\kappa}_1(t)=(1+f^2(t))\kappa_1^2(t)+2\kappa_1(t)\frac{\dot{\gamma}_1(t;n_1)}{\gamma_1(t;n_1)}.
\end{align}
Hence, we can compute $\kappa_1(t)$, and consequently, the rest of the controllable decay rates. Having obtained this expression, it is straightforward to extend it to $N$ QCCs, that is,
\begin{align}\label{eq:kappaQWVN}
    \dot{\kappa}_1(t)=\left[\sum_{j=1}^N f_j^2(t)\right]\kappa_1^2(t)+2\kappa_1(t)\frac{\dot{\gamma}_1(t;n_1)}{\gamma_1(t;n_1)}.
\end{align}
where $f_j(t)=\gamma_j(t;n_j)/\gamma_1(t;n_1)$ so that $f_1(t)=1$ by definition. Clearly, for $N=1$, one recovers Eq.~\eqref{eq:kappaQWV}.

For example, let us consider $N=2$ QCCs for sech-photons as $\gamma_j(t;n_j)=\sqrt{\kappa_{p,j}/(4n_j)}{\rm sech}(\kappa_{p,j} t/2)$ (same time of emission), so that $f_j(t)=\sqrt{\kappa_{p,j}n_1/(\kappa_{p,1}n_j)}\cosh(\kappa_{p,1} t/2){\rm sech}(\kappa_{p,j}t/2)$. Note that there is no need to consider simultaneous emission, but it simplifies the expressions. One can solve Eq.~\eqref{eq:kappaQWVN} imposing as well that $q(t)=\gamma_1(t;n_1)/\sqrt{\kappa_1(t)}$ fulfills $\lim_{t\rightarrow-\infty} |q(t)|^2=1$ and $\lim_{t\rightarrow\infty} |q(t)|^2=1-\frac{1}{n_1}-\frac{1}{n_2}$, 
\begin{align}
\kappa_1(t)= \frac{\kappa_{p,1}n_2(1+\cosh(\kappa_{p,2}t)){\rm sech}^2(\kappa_{p,1} t/2)}{2(n_1+n_2-2n_1n_2+(n_1+n_2-2n_1n_2)\cosh(\kappa_{p,2}t)+n_1\sinh(\kappa_{p,2}t)+2n_2\cosh^2(\kappa_{p,2} t/2)\tanh(\kappa_{p,1}t/2)}
\end{align}
while $\kappa_2(t)=f_2^2(t)\kappa_1(t)$. Note that $n_1+n_2\geq 1$. 

For a general case of $N$ QCCs, but restricting to $\kappa_p\equiv \kappa_{p,j}$ and no delay among emitted photons, one finds a rather simple expression, 
\begin{align}
    \kappa_1(t)=\frac{k_p{\rm sech}^2(\kappa_p t/2)}{n_1(4-2\tilde{n}-2\tilde{n}\tanh(\kappa_p t/2))}
\end{align}
with $\tilde{n}=\sum_{j=1}^Nn_j^{-1}$.

\section{Genuine multipartite entanglement: Entanglement of formation}
Besides the fidelity of the generated state, we quantify genuine multipartite entanglement via the entanglement of formation~\cite{Wootters98,Szalay15}. For a tripartite system and a pure state $\rho$, we have
\begin{align}
E_{\rm 3F}(ABC)\equiv \min\left\{S(\rho_A),S(\rho_B),S(\rho_C) \right\}
  \end{align}
where $S(\rho)=-{\rm Tr}[\rho\log_2\rho]$ is the von Neumann entropy, and $\rho_{x}$ refers to the reduced states for $x=A,B,C$ subsystems. In general, for mixed states $\rho$, the entanglement of formation follows from a minimization over all possible pure-state decomposition of $\rho$, namely,
\begin{align}\label{eq:e3fmixed}
    E_{\rm 3F}(ABC)\equiv \min_{\ket{\psi_i}}\left( \sum_i p_i \min \{S(\rho_{A,i}),S(\rho_{B,i}),S(\rho_{C,i}) \}\right)
\end{align}
with $\rho=\sum_i p_i \ket{\psi_i}\bra{\psi_i} $ and $\rho_{x,i}$ denotes the reduced state for the $x$ qubit for a global pure state $\ket{\psi_i}$. 
The tripartite entanglement of formation $E_{3F}$ is invariant under local operations and non-increasing under LOCC, and $E_{3F}>0$ if and only if $\rho$ is genuinely tripartite entangled. 
In order to compute the minimization, we proceed following~\cite{Wootters98}, i.e. we compute any ensemble that gives rise to $\rho$  using a unitary matrix $U$ of dimension $M\times M$ with $M\geq n$ where $n$ denotes the rank of $\rho$, i.e. $\rho=\sum_{i=1}^n \lambda_i \ket{v_i}\bra{v_i}$ its spectral decomposition. Then, one can write an ensemble $\{\ket{\omega_k}\}$ with $k=1,\ldots, M$ such that $\rho=\sum_{k=1}^M \ket{\omega_k}\bra{\omega_k}$ where
\begin{align}
    \ket{\omega_k}=\sum_{j=1}^n U_{i,j}\sqrt{\lambda_j} |v_j\rangle.
\end{align}
This allows for the minimization in Eq.~\eqref{eq:e3fmixed}.

\end{document}